\documentclass{JAC2003}

\usepackage{graphicx}
\usepackage{tabularx}
\usepackage{amssymb}

\begin{document}
\title{Summary of SLAC'S SEY Measurement On Flat Accelerator Wall Materials
\thanks{Work supported by the U.S. Department
of Energy under contract number DE-AC02-76SF00515}}

\author{F. Le Pimpec \\ Paul Scherrer Institute, 5232 Villigen
Switzerland \\R.E. Kirby, F.K. King, and M. Pivi \\
Stanford Linear Accelerator Center, Menlo Park, CA 94025 \\}

\maketitle

\begin{abstract}

The electron cloud effect (ECE) causes beam instabilities in
accelerator structures with intense positively charged bunched
beams. Reduction of the secondary electron yield (SEY) of the beam
pipe inner wall is effective in controlling cloud formation. We
summarize SEY results obtained from flat TiN, TiZrV and Al
surfaces carried out in a laboratory environment. SEY was measured
after thermal conditioning, as well as after low energy, less than
300~eV, particle exposure.

\end{abstract}

\newcommand{\registered}{\circledR}

\section{Introduction}
Multipactoring and the ECE have been detrimental to the
functioning of onboard space devices as well as klystrons and
accelerators. In the latter, the ECE was characterized and dealt
with at the CERN ISR (Intersecting Storage Rings)
\cite{Calder:1973,grobner:1977}. With the construction of high
current colliders, e.g., the B factories, or accelerators
producing a high amount of photons per (e$^+$, H$^+$) bunch,
multipacting and the ECE are again being studied extensively
\cite{Pivi:pac05}.

Methods to suppress multipacting or the formation of electron
cloud (EC) are still similar to the ones used for radio frequency
(RF) components. One can allow the secondary electron to be
produced and then get rid of them afterwards. For example by the
use of electron clearing electrodes or, by use of low magnetic
field solenoids that return the secondary electrons to the
surfaces from which they have been produced. When photons are
responsible for the creation of the EC, it is also possible to
confine the photoelectrons to a place which is non-detrimental to
the accelerator, for example in an ante-chamber.

Finally, one may modify the surface to produce less than one
secondary  electron  per  incident electron. This can be done, for
example, by using a rough surface, by using an emission
suppressing coating, or by cleaning the surface in-situ (thermal
treatment, plasma glow discharge etc...) to remove high SEY
adsorbed gas and oxide layers. These remedies can also be mixed
together to give the best results for solving the problem in an
existing machine or to be applied in a forthcoming accelerator
where the ECE problem is expected to occur. Of course, lowering
the circulating beam intensity can retard cloud formation but that
also affects luminosity. In this summary, we report SEY
measurements obtained at SLAC on flat surfaces, aluminium, TiN,
TiZrV Non Evaporable Getter (NEG) and TiCN.

\section{SEY experimental setup}

\begin{figure}[tbph]
\begin{center}
\includegraphics[width=0.6\textwidth,clip=]{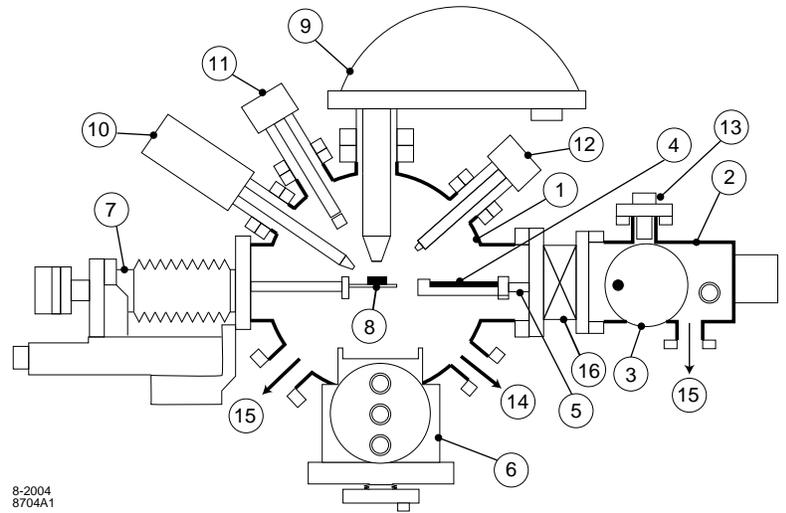}
\end{center}
\caption{Experimental ultra high vacuum (UHV) system.}
\label{figsketchsetup}
\end{figure}

The system used to measure the SEY is shown in
Fig.\ref{figsketchsetup}. The experimental  methodology  used  to
measure  the  secondary electron yield has been described in
\cite{lepimpec:jvsta2005}

\begin{Itemize}
\item 1 Analysis chamber - 2 Loadlock chamber
\item 3 Sample plate entry - 4 Sample transfer plate
\item 5 Rack and pinion travel - 6 Sample plate stage
\item 7 XYZ$\theta$ Omniax$^{TM}$ manipulator - 8 Sample on XYZ$\theta$
\item 9 Electrostatic energy analyzer - 10 X-ray source
\item 11 SEY/SEM electron gun - 12 Microfocus ion gun
\item 13 Sputter ion gun - 14 To pressure gauges and RGA
\item 15 To vacuum pumps - 16 Gate valve
\end{Itemize}

The SEY ($\delta$) definition is the number of electrons leaving
the surface over the number of incident electrons (primary
electrons), which becomes $\delta = 1 - {I_T}/{I_P}$. With ${I_P}$
the primary electrons and ${I_T}$ the total sample current being
the difference between the primary and secondary electron current.

\section{Conditioning}

\subsection{Thermal conditioning}

A natural method for reducing the SEY of a material is thermal
heating. Usually this is achieved during an in-situ bakeout.
However, to be efficient the bake should be above 150$^\circ$C.
Nevertheless, any increase of surface temperature has an effect on
the SEY \cite{Calder:1986,benven:1998}. On some materials like Cu
or Ag, certain oxides have an SEY below that of the atomically
clean metal. Hence, growing an in-situ oxide by heating the
surface, in presence of oxygen, is also a possibility
\cite{Scheuerlein:2003}. Many examples can be found in the
literature, and an excellent summary is available
\cite{CERNecloudweb}.

\subsection{Particle conditioning}

Another way of processing a surface to lower its SEY is to expose
it to energetic particles :  photons, electrons or ions. Usually
the SEY of metals obtained after exposure to energetic particles
is close to that of an atomically clean surface. This trend seems
also not to be observed in the case of exposure to very energetic
ions, MeV range per nucleons \cite{Hanson:2001}.

In the laboratory we have quantified the reduction of the SEY as a
function of electron or ion bombardment. By ion bombardment we
mean an ion beam, not a plasma glow discharge. A plasma glow
discharge is very effective in cleaning the surface in a few
minutes, but plasma gas pressure required for a stable discharge
is above a mTorr \cite{Calder:1986}. Moreover, performing an
in-situ glow discharge of in an accelerator vacuum beam pipe is
far from trivial.

\section{Conditioning of Ti-Based Coatings}

TiN and activated TiZrV getter coatings are good candidates for
suppressing the ECE. TiN coating is known to have a SEY max
($\delta_{max}$) below or close to 1 when freshly deposited
\cite{Garwin:1987,kirby:2001}. However, when the "as-deposited"
film is  exposed  to  air, its  SEY maximum  varies  between  1.5
to  2.7 \cite{he:EPAC04,henrist:9803}. Sputtering of air exposed
TiN by Ar$^+$ ions or exposing it to a high dose of electrons will
return $\delta_{max}$ to around 1
\cite{kirby:2001,he:EPAC04,Prieto:1995,Kato:AVS2002}.

\begin{figure}[tbp]
\centering
\includegraphics[width=0.55\textwidth,clip=]{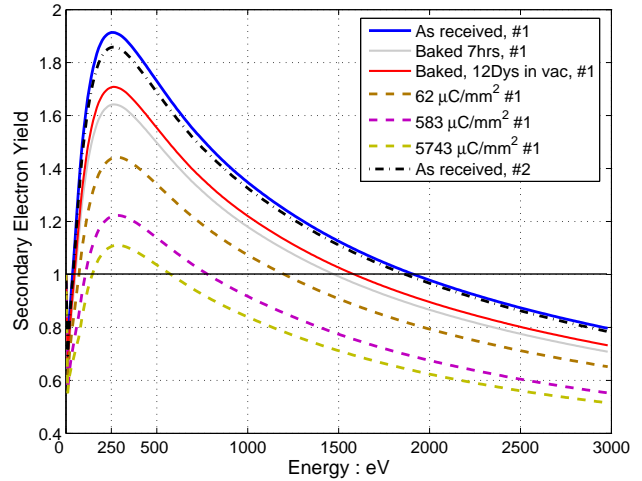}
\caption{SEY of TiN/Al under different conditions.
As-received (\#1 and \#2),  baked at 150$^\circ$C, vacuum recontamination after
12 days at 5.10$^{-10}$ Torr and conditioning by 130~eV electrons.
Measurement performed at 23$^\circ$ primary incidence.}
\label{figSEYLBLall}
\end{figure}

\begin{figure}[tbp]
\centering
\includegraphics[width=0.55\textwidth,clip=]{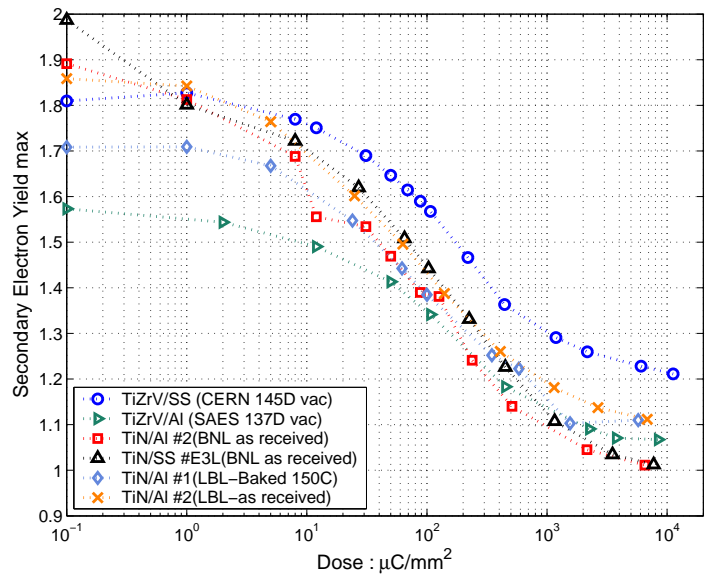}
\caption{SEY of TiN and TiZrV getter under exposure by 130~eV
electrons. Measurement performed at 23$^\circ$ primary incidence.}
\label{figSEYmaxelectron}
\end{figure}

In the ECE the energy gain of the secondary electrons is typically
lower than 300~eV. In the case of the International Linear
Collider (ILC) positron damping ring, the average energy was
computed to be 130 eV \cite{lepimpec:Nima2005}. The effect of
130~eV electron conditioning  on the SEY and $\delta_{max}$ are
shown in Fig.\ref{figSEYLBLall} and \ref{figSEYmaxelectron},
respectively.

\begin{figure}[tbp]
\centering
\includegraphics[width=0.52\textwidth,clip=]{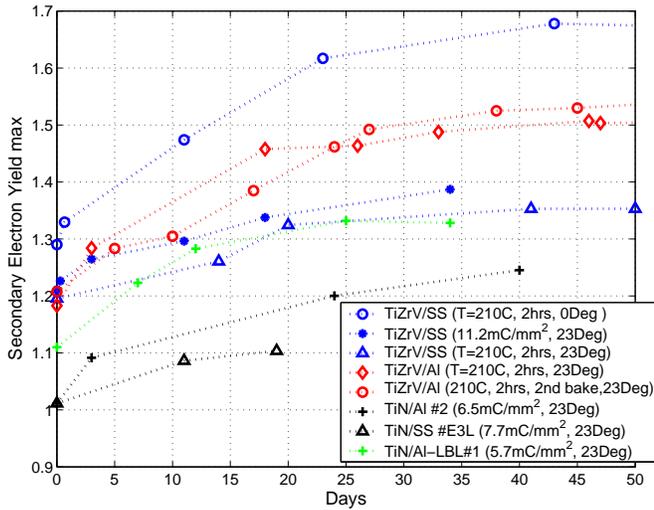}
\caption{TiN and TiZrV NEG SEY increase when left in baked UHV
atmosphere of 5.10$^{-10}$ Torr.}
\label{figSEYmaxvacuum}
\end{figure}

It can be seen that after a surface has been cleaned by a bake
(Fig.\ref{figSEYLBLall}) or by electron conditioning the SEY will
increase even when the surface is left under a good vacuum
(5.10$^{-10}$ Torr), Fig.\ref{figSEYmaxvacuum}. Any scrubbed or
"clean" surface will adsorb molecules from the residual gas. The
increase of the SEY is linked to the oxidation of the surface by
the presence of oxygen atoms in these molecules. This was directly
verified by observing the evolution  of  the XPS (x-ray
photoelectron spectroscopy) carbon spectrum of TiZrV during
exposure to residual gas \cite{lepimpec:Nima2005}.

During the passage of the circulating positron beam, ions will be
created from the residual gas. Their energy, for the ILC damping
rings, is less than 200~eV. Bombardment from these ions can also
contribute to SEY reduction. To probe this effect, we have
submitted three surfaces to ion bombardment. The results are
summarized in Table.\ref{tabSEYiondose}. The experimental
parameters are described in \cite{lepimpec:Nima2006}.

\begin{table}[tbph]
\centering \caption{$\delta_{max}$ reduction due to 250~eV ion
conditioning. SEY measured at normal incidence}
\begin{tabularx}{\linewidth}{|X|X|c|c|X|}
  \hline
  Thin Film & ion species & $\delta_{max}$ & Energy$_{max}$ & Dose $\mu C/mm^2$ \\
  \hline
  TiCN & H$_2$ & 1.29 & 280 & 1.11 \\
  TiN & N$_2$ & 1.09 & 260 & 3.39 \\
  TiZrV & N$_2$ & 1.15 & 300 & 2.29 \\
  \hline
\end{tabularx}
\label{tabSEYiondose}
\end{table}

Comparing the effect of conditioning to TiN, a gas-saturated TiZrV
NEG was conditioned with a 130~eV electron beam,
Fig.\ref{figSEYmaxelectron}, and to a 250~eV ion beam,
Table.\ref{tabSEYiondose}. An N$_2 ^+$  ion dose of
0.96~$\mu$C/mm$^2$ reduces $\delta_{max}$ from 1.45 to 1.18,
further exposure up to 2.29~$\mu$C/mm$^2$, causes only a
$\delta_{max}$ decrease from 1.18 to 1.15
\cite{lepimpec:Nima2006}.

\section{Technical Aluminium under electron exposure}

Aluminium is one of the common metal used in fabricating the
accelerator vacuum chambers. The SEY of atomically clean Al ranks
among the best material with a $\delta_{max}$ around 1. Clean Al
is extremely reactive to oxygen, however, upon air exposure, it
will form a thick oxide with a $\delta_{max}$ well above 2
\cite{lepimpec:jvsta2005}. As discussed earlier, electron
conditioning will bring the $\delta_{max}$ of the metal to its
atomically clean value.

\begin{figure}[htbp]
\centering
\includegraphics[width=0.52\textwidth,clip=]{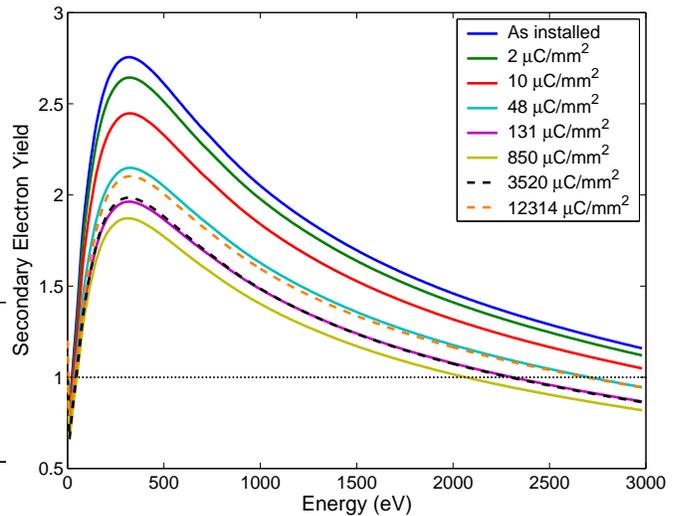}
\caption{Al 1100 exposed to electron conditioning. The primary
electron beam was impinging at 23$^\circ$ from normal incidence.}
\label{figSEYAlu}
\end{figure}

We repeated that measurement with oxidized aluminium. In the
laboratory, we conditioned three different air-exposed technical
aluminium surfaces and observed that the SEY decreases at first
and then re-increases. Results shown in Fig.\ref{figSEYAlu} are
similar for the two other samples \cite{lepimpec:jvsta2005}.

During conditioning the XPS spectra show the C1s peak shifting
toward lower binding energy (BE)), signaling reduction of the
oxide surface. With further conditioning, the trend stops and the
peak broadens. Atomically clean Al shows one peak at 73~eV
(metallic) and another, Al$_2$O$_3$, at 76~eV. Two of our samples
show this double peak structure. Thus, during conditioning, the
peaks evolve from oxide to clean Al and then reverse again. The
third sample was extremely oxidized but, again, the broad peak
shifted to lower BE and then broadened further, consistent with
the other samples \cite{lepimpec:jvsta2005}.

\section{NEG pumping and SEY}

Activated getter surfaces \cite{Meli:1990,benven:99}, like
St707$^{\registered}$, TiZr and TiZrV have a $\delta_{max}$ below
1.3 \cite{Bojko:1998,Henrist:2001}. They also provide
linearly-distributed pumping capacity. The main interest in TiZrV
coating over the other NEG alloys is that its activation
temperature is lowest, 180$^\circ$C \cite{Chiggiato:2006}. During
residual gas pumping the SEY of NEG increases
(Fig.\ref{figSEYmaxvacuum}), just as it does for initially-clean
TiN. The SEY increase of NEG was followed with XPS measurement by
monitoring the evolution of the carbon peak
\cite{lepimpec:Nima2005}. Significantly, during residual gas
atmosphere saturation of NEG, the $\delta_{max}$ exceeds 1.3, the
maximum value obtained when saturating NEG with individual common
residual gases present in an accelerator environment
\cite{Henrist:2001}. This behaviour is seen for a freshly
activated NEG as well as from a 11.2~mC/mm$^2$ electron
conditioned surface, Fig.\ref{figSEYmaxvacuum}. This suggests that
fast saturation by a single species is different from slow
saturation by multiple species, i.e., that time may play a role or
co-adsorption of multiple species, may enhance the surface
oxidation mechanism, as it can be seen in some surfaces
\cite{Gupta:1997}. The residual gas composition of a baked UHV is
mainly composed of H$_2$ which readily diffuses into the NEG, CO,
CO$_2$ , H$_2$O, and CH$_4$ which is negligibly-pumped by TiZrV
\cite{Chiggiato:2006}. The co-adsorption of the three oxidizing
species enhances oxidation, similar to an air-oxidation, compared
to oxides formed by dosing with a single specie. Thicker surface
oxide has generally higher SEY. The process involved in building
this thick oxide might be somewhat equivalent to cryogenic
co-adsorption \cite{wallen:1996}. It is planned to test this
hypothesis in our setup.

\section{Conclusions}

We have investigated a series of flat surface materials for
suppressing the ECE. The most promising remedies are Ti-based
coatings. TiN has historically been the choice for successfully
reducing multipacting. Upon conditioning exposure to low energy
ions or electrons, its atmosphere-oxidized surface $\delta_{max}$
returns to 1. TiCN was developed as an oxidation-resistant
replacement for TiN;  however, it behaves similarly to TiN.
Another coating option is low temperature activated NEG, TiZrV.
Following an activating-bake, its $\delta_{max}$ is 1.2. As a
bonus, NEG coating provides distributed beam chamber wall pumping.
While pumping toward film gas saturation, the $\delta_{max}$
increases. To return the SEY to low value and restore the pumping
capacity, the film can be thermally re-activated multiple times.
The SEY of the surface itself may also be restored by electron or
ion bombardment, which will also recreate some surface pumping
capacity.

Technical Al surfaces were investigated under electron exposure.
We found that $\delta_{max}$ will not go consistently below 1.8.
However, the behaviour at very large doses, above 3.10$^4$~$\mu
C/mm^2$, was not measured. The SEY may increase further, stabilize
or oscillate.

In an accelerator environment, synchrotron radiation, ions and
electrons not only desorb molecules but also produce electrons
which can lead to the formation of the EC. As we have seen,
electron exposure is very efficient in reducing the SEY. However,
as the EC conditions the surface, the number of secondary
electrons diminishes, hence the EC can oscillate between being ON
or OFF. Nevertheless, photons and ions directly created by the
beam may ensure that an EC does not develop, but only direct
measurements in beam chambers will confirm this
\cite{Suetsugu:Nima2006,Pivi:ecloud07}.

\section{Acknowledgments}
Most of the thin film sample were graciously produced by D. Lee
and A. Wolski at LBNL, P. He and H.C Hseuh at BNL. Our first TiZrV
getter was produced at CERN, thanks to V. Ruzinov of the EST
group.

%
%


\end{document}